\documentclass[a4paper,twoside]{article}
\newcommand{\affil}[1]{$^{\rm #1}$}
\usepackage[authoryear]{natbib}
\bibpunct{(}{)}{;}{a}{}{,}
\usepackage{graphicx}
\newcommand{\ox}[1]{\mbox{$^{#1}$O}}
\newcommand{\fluor}[1]{\mbox{$^{#1}$F}}

\newcommand{\al}[1]{\mbox{$^{#1}$Al}}

\def\araa{Ann. Rev. Astron. Astrophys.}
\def\apj{Astrophys. J.}
\def\apjl{Astrophys. J.}

\def\aap{Astron. Astrophys.}
\def\mnras{MNRAS}

\def\prc{Phys. Rev. C}
\def\prl{Phys. Rev. Lett. }

\date{} 
\title{\large\bf\flushleft On the Mass and Metallicity Distributions of the Parent AGB Stars of O-rich Presolar Stardust Grains}
\author{\parbox{\textwidth}{\flushleft
\vspace{-0.5cm}
{\it Larry R. Nittler\affil{A,B}}\\
\vspace{0.4cm}
{\small \affil{A}\,Dept.\ of Terrestrial Magnetism, Carnegie Institution of Washington, Washington DC, 20015, USA}\\
{\small \affil{B}\,Email: lnittler@ciw.edu}}}
\begin{document}
\twocolumn[
\begin{changemargin}{.8cm}{.5cm}
\begin{minipage}{.9\textwidth}
\vspace{-1cm}
\maketitle
%
%
\small{\bf Abstract:}
Presolar grains in meteorites formed in a sample of AGB stars that ended their lives  within $\approx$1 Gyr of the origin of the Solar System 4.6 Gyr ago. The O-isotopic compositions of presolar O-rich stardust reflect the masses and metallicities of their parent stars. We present simple Monte Carlo simulations of the parent AGB stars of presolar grains. Comparison of model predictions with the grain data  allow some broad conclusions to be drawn: 1) Presolar O-rich grains formed in AGB stars of mass $\sim$1.15 -- 2.2 M$_{\odot}$. The upper-mass cutoff reflects dredge-up of C in more massive AGB stars, leading to C-rich dust rather than O-rich, but the lack of grains from intermediate-mass AGB stars ($>$4M$_{\odot}$) is a major puzzle. 2) The grain O-isotopic data are reproduced well if the Galaxy in presolar times was assumed to have a moderate age-metallicity relationship, but with significant metallicity scatter for stars born at the same time. 3) The Sun appears to have a moderately low  metallicity for its age and/or unusual \ox{17}/\ox{16} and \ox{18}/\ox{16} ratios for its metallicity. 4) The Solar \ox{17}/\ox{18} ratio, while unusual relative to present-day molecular clouds and protostars, was not atypical for the presolar disk and does not require self-pollution of the protosolar molecular cloud by supernova ejecta.


\medskip
\medskip
\end{minipage}
\end{changemargin}
]
\small

\section{Introduction}

Primitive extraterrestrial materials (e.g., meteorites, interplanetary dust particles) contain a small amount of ``presolar grains,''  tiny dust grains which formed in previous generations of stellar outflows and supernova explosions \citep{Nittler2003,Zinner2007,Clayton2004}.  These grains retain the isotopic compositions of the stellar gases from which they condensed and have proven a very valuable source of information for an array of astrophysical processes. The isotopic, elemental and mineralogical compositions of presolar grains reflect a complex interplay of galactic chemical evolution (GCE), nucleosynthesis, and the physicochemical conditions of stellar dust formation. 

A wide variety of presolar phases has been identified, including carbides, oxides, silicates and nitrides. Here we  consider O-rich stardust grains, comprising a variety of oxide and silicate phases, as these are the most abundant unambiguously presolar grains in meteorites, and O-rich dust dominates the interstellar dust budget. A large majority of presolar oxides and silicates is believed to have formed in O-rich asymptotic giant branch (AGB) stars (\S 2). Because the typical lifetime of stellar grains in the interstellar medium (ISM) is  estimated to be $\approx$0.5 Gyr \citep{Jones96}, the parent stars of presolar grains must have ended their lives relatively shortly before the Solar System formed 4.6 Gyr ago.   However, because stellar lifetimes depend strongly on mass and the grains formed in stars with a range of masses, the parent stars must have formed over a long history of the galactic disk \citep{Nittler97}. The grains thus represent a sample of stars that were present billions of years ago and form a complementary dataset to astronomical observations for studies of GCE. In this paper, we present simple Monte Carlo models designed to investigate the parent mass and metallicity distributions of presolar grains. Although oversimplified, our approach is valuable for estimating  the mass range of AGB stars contributing O-rich dust to the ISM. Moreover, the grains can shed light on important questions in GCE, including the existence of an age-metallicity relationship in the Milky Way disk and the \ox{17}/\ox{18} ratio of the Galaxy in presolar times.  

\section{O-rich Presolar Stardust}

\begin{figure}[h!]
\begin{center}
\includegraphics[scale=.65, angle=0]{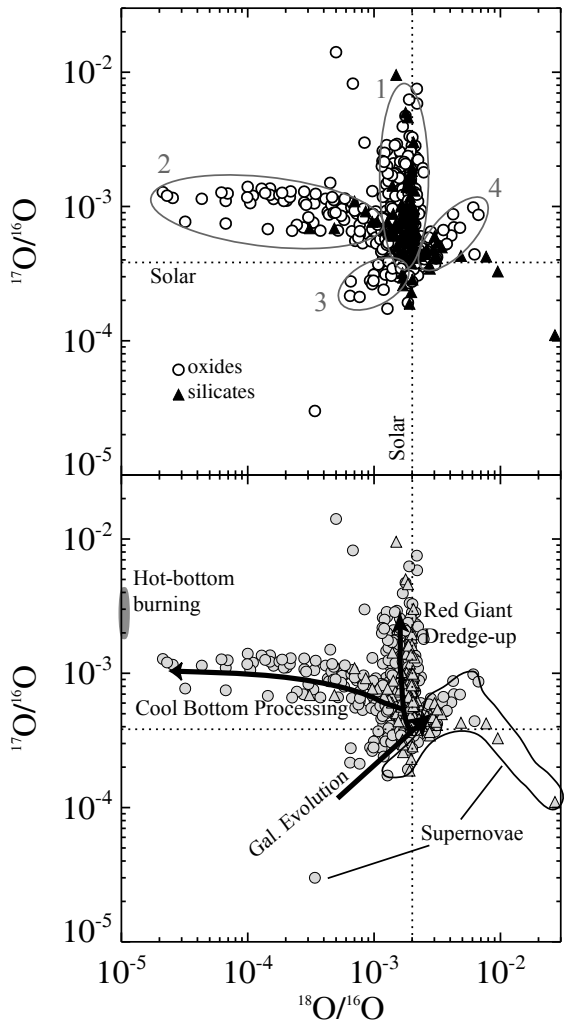}
\caption{O isotopes in presolar oxide and silicate grains; see \citet{Nittler2008} for data sources. Ellipses in the top panel indicate grain Groups defined by \protect\citet{Nittler97b}.  Expected trends for different stellar sources and/or processes are shown in the bottom panel.}\label{oxfig1}
\end{center}
\end{figure}

Fig.~\ref{oxfig1} shows the O-isotopic ratios for several hundred presolar oxide and silicate grains, divided into Groups according to  \citet{Nittler97b}. The vast majority of grains have \ox{17} excesses and slight to strong \ox{18} depletions, characteristic of H-burning by the CNO cycles. As indicated, the dominant Group 1 population is well-explained by models of the first dredge-up, which occurs as all stars become red giants and mix ashes of partial H-burning into their convective envelopes. Comparison of the Group 1 isotopic signatures with observations of O-rich AGB stars, as well as the inferred initial presence of \al{26} in many of the grains, supports an origin in AGB stars \citep{Nittler97b}. The predicted O-isotopic composition of an AGB star envelope  depends on the mass and initial composition of the star, as well as any extra mixing (``cool bottom processing'')  that occurs \citep{Boothroyd94,Boothroyd99,Wasserburg95a}. The \ox{17}/\ox{16} ratio is primarily set by the first dredge-up and depends strongly on stellar mass with only a minor dependence on  the initial composition. In contrast, the \ox{18}/\ox{16} ratio is affected only slightly (up to 20\%) by the first dredge-up, so that larger variations must reflect variations in initial composition and/or additional processing through cool bottom processing. The latter is believed to explain the large \ox{18} depletions observed in the Group 2 presolar grains, though the physical cause of the mixing is not yet well understood \citep{Nollett2003,Nordhaus2008}. As discussed earlier \citep{Nittler97b,Nittler97} and below, the slightly \ox{16}-rich Group 3 grains can be explained as having formed in AGB stars of low mass ($<$1.2M$_{\odot}$) and metallicity, since lower-metallicity stars are expected to have formed with lower abundances of the secondary isotopes \ox{17} and \ox{18} and the first dredge-up only has a minor impact on the surface O-isotopic composition in such low-mass stars. Previous work \citep{Nittler97b} showed that the general shape of the Groups 1 and 3 O-isotopic distribution could be easily explained by a simple combination of GCE and dredge-up models. The work presented here provides further elaboration of this point, but moves beyond it to take into account the density of data points on the O 3-isotope plot.

The origin of the  \ox{18}-enriched Group 4 grains is more ambiguous than the other grains. However, multi-element isotopic compositions of a few Group 4  oxide \citep{Nittler2008}  and silicate \citep{Vollmer2008} grains strongly support a supernova origin for the majority of these grains. A supernova origin is also most likely for two  strongly-\ox{17}-depleted grains as well \citep[Fig.~\ref{oxfig1}]{Nittler98b,Messenger2005}.

\section{Monte Carlo Simulations}

The distribution of O isotopes in the presolar grains indicates that the parent stars had a range of stellar masses and initial isotopic compositions, most likely related to metallicity \citep{Nittler97b}. A key question is how the distributions of masses and metallicities inferred for the grains compare to what might be expected for a population of AGB stars that would have contributed dust to the ISM close in time and space to the formation of the Solar System. A detailed treatment of this problem would require coupling of a full GCE model to models of dust production in stars of different mass and metallicity  and destruction in the ISM  (see, e.g., Gail et al. 2008)\nocite{Gail2008}. The approach used here is greatly oversimplified yet still useful for addressing some basic issues.  Monte Carlo techniques are used to generate synthetic mass and metallicity  distributions for a set of stars and the predicted O isotopic distributions (based on interpolation of first dredge-up calculations) are compared to the observed grain distributions. Note that although close to one thousand presolar oxide and silicate grains have now been identified, most have been found with techniques that could  bias or compromise the dataset. For example, all silicates have been found in situ in planetary materials or on very crowded sample mounts where there is likely to be contribution to the O isotopic measurements from surrounding normal material, due to tails on the primary ion beam in the ion microprobe \citep{Nguyen2007}.  To have the most unbiased sample,  we  restrict consideration to the $\sim$200 grains that have been identified by measurements of all three O isotopes in well-separated grains. The simulations thus generate 200 stars with masses and metallicities taken randomly from the distributions described below. 

{\it Mass Distributions}. A critical component of  GCE studies is the initial mass function (IMF), which describes the distribution of masses for a single stellar generation.  What is needed for comparison to presolar grain data is the ``solar time mass function (STMF),'' the distribution of masses of stars ending their lives shortly before Solar System formation. The shape of the STMF at a given time depends on the whole history of star formation prior to that time. Because of the longer lives of lower mass stars compared to higher mass ones, the STMF starts out top-heavy in the early Galaxy but will gradually move down to include more and more lower-mass stars as sufficient time passes for earlier generations of lower-mass stars to live out their lives. The lower mass limit of the STMF is set by the age of the Galaxy. For example, assuming that the Galaxy was 8.5 Gyr old at the time of solar birth indicates that no stars with lifetimes longer than this (mass about 1.15 M$_{\odot}$)  could possibly contribute presolar grains to the Solar System. For simplicity, we assume that by the time of solar birth, GCE had reached a quasi-steady state such that the shape of the STMF approximated that of the IMF, except for the low-mass cutoff due to the age of the Galaxy. This is borne out by the detailed modeling of \citet{Gail2008}.  For the Monte Carlo simulations, stellar masses are  taken from the usual \citet{Salpeter55} IMF, where the number  of stars varies as M$^{-2.35}$.

{\it Metallicity Distributions}. For each simulated star, the metallicity is chosen from a Gaussian distribution, where the width of the distribution represents the intrinsic  metallicity spread for stars born at the same time \citep{Edvardssen93,Reddy2003}. This width, $\sigma_{Z}$, is assumed to be constant in time in terms of absolute value and is specified by its relative value at solar metallicity. The average metallicity of the distribution depends on the stellar mass and an assumed age-metallicity relationship (AMR), reflecting the expectation that stars born more recently have, on average, higher metallicity than stars that were born earlier.  For simplicity, the average metallicity, expressed as the logarithmic iron abundance, is assumed to vary linearly with time. Explicitly, each star is assumed to have formed at a time $\tau$ before the formation of the Solar System, given by a  stellar mass-lifetime relation \citep{Mathews92}. The average [Fe/H]  for stars of mass M is then given by the expression: ${\rm [Fe/H]}= a - b \cdot \tau({\rm M})$, where  $a$ is the average metallicity [Fe/H] at the time of Solar birth and $b$ is the slope of the age-metallicity relationship. For the present limited study, we consider two values of $b$: $b=$0.02 dex/Gyr, which is consistent with the age-metallicity distribution derived by \citet{Haywood2006} and $b=0$, representing the limiting case of there being no AMR in the late Galaxy \citep{Nordstrom2004}. Because the absolute value of the metallicity spread $\sigma_{Z}$ is constant in time, its relative value increases with decreasing metallicity. 

{\it Oxygen Isotopes}. We use the first (and second) dredge-up calculations of \citet[hereafter BS99]{Boothroyd99} to predict the O isotopic compositions of the simulated stars. These authors calculated the surface CNO isotopes for a grid of stellar masses (M=0.85 -- 9M$_{\odot}$) and metallicities, Z, ranging from $10^{-4}$ to 0.02. The initial O isotopic ratios were assumed to vary linearly with metallicity, since \ox{16} is a primary isotope and \ox{17,18} are secondary. The precise evolution of O isotopes in the Galaxy is highly uncertain, however \citep{Prantzos96}, so for the present study, we have assumed various relationships between the initial \ox{17}/\ox{16} and \ox{18}/\ox{16} ratios and metallicity. For each simulated star, we then interpolate or extrapolate the BS99 models to predict a post-dredge-up surface composition for the chosen mass and initial composition.

The basic parameters to be varied in our simple Monte Carlo model are the minimum and maximum masses of the STMF, the AMR parameters $a$ and $b$ described above, the width, $\sigma_{Z}$ of the metallicity distribution for stars born at a given time, and the assumed initial O isotope-metallicity relationship. Note that these are not all independent from each other. In particular, the \ox{18}/\ox{16} ratio of an AGB star envelope  depends strongly on the stellar initial composition, which in the present model is based both on the metallicity and the assumed relationship between initial O isotopic ratios and metallicity. It is thus  not possible to find  a unique set of parameters that best matches the presolar grain data. That said, the main goal of this work is to explore what broad conclusions might be drawn from  the presolar grain data, not to find absolute quantitative values for specific model  parameters.

\section{Mass Distributions}

\begin{figure}[h]
\begin{center}
\includegraphics[scale=.5, angle=0]{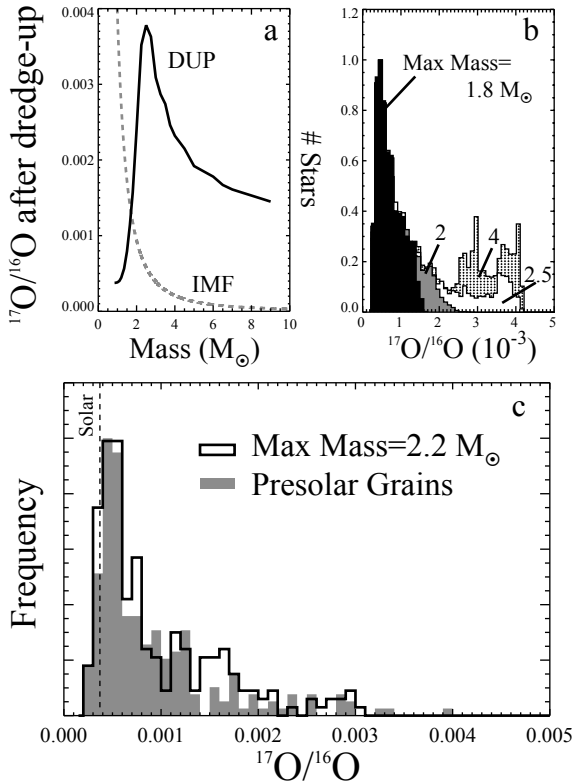}
\caption{(a) Predicted  \ox{17}/\ox{16} ratios in the envelopes of AGB stars following the first dredge-up (``DUP,'' Boothroyd \& Sackmann 1999)   and the assumed initial mass function (``IMF,'' Salpeter 1955) as a function of stellar mass. (b) Histograms of \ox{17}/\ox{16} ratios for simulated distributions of stars with a minimum mass of 1.15 M$_{\odot}$ and 4 indicated maximum masses. (c) Comparison of simulated and observed \ox{17}/\ox{16} ratios for Monte Carlo model with a maximum mass of 2.2 M$_{\odot}$. }\label{massfig}
\end{center}
\end{figure}

Let us first consider the masses of potential presolar grain parent stars. As stated earlier, the \ox{17}/\ox{16} ratio of an AGB stellar envelope is  strongly dependent on the mass of the star. As a result, the distribution of \ox{17}/\ox{16} ratios in presolar AGB grains is a sensitive measure of the mass distribution of the parent stars. This is illustrated by Figure \ref{massfig}. The predicted \ox{17}/\ox{16} ratio following the first dredge-up increases with stellar mass up to $\sim$2.5M$_{\odot}$ and then decreases again for higher masses (``DUP'' in 
Fig.~\ref{massfig}a). As a result, if a continuum of stellar masses spanning the peak in \ox{17}/\ox{16} ratios produced grains, one might expect a pile-up of compositions near the peak value of   \ox{17}/\ox{16}$\approx$0.003 -- 0.004.  Fig.~\ref{massfig}b shows histograms of predicted \ox{17}/\ox{16} ratios for Monte Carlo simulations with a Salpeter IMF with a minimum mass of 1.15 M$_{\odot}$ and  four different maximum masses (1.8 -- 4 M$_{\odot}$). Clearly, the number of stars with  \ox{17}/\ox{16}$>$0.002 depends critically on the maximum mass allowed in the IMF in the model. In Fig.~\ref{massfig}c we compare the results of the model with a mass range of  1.15 --  2.2 M$_{\odot}$ to the distribution of Groups 1--3 presolar oxide grains. Note that we include Group 2 grains in the histogram because the extra-mixing thought to be responsible for their \ox{18} depletions is expected to have a much smaller effect on their \ox{17}/\ox{16} ratios and the latter thus should mostly  reflect the first dredge-up as it does for Group 1 grains. In any case, the Monte Carlo distribution compares well with the observed one and the relative lack of grains with  \ox{17}/\ox{16}$>$0.002 indicates that there were few parent stars with masses greater than $\sim$2.2 M$_{\odot}$. In particular, there is no evidence for any significant  fraction of grains from intermediate-mass AGB stars (e.g. 4--7 M$_{\odot}$) as these would form a large (not observed) peak in the histogram around 0.002--0.004. Note also that simulations extending to lower masses than 1.15 M$_{\odot}$ over-predict compositions at the low end of the \ox{17}/\ox{16} distribution.

The lower limit on the inferred mass range of 1.15 M$_{\odot}$ can be easily understood,  as explained above, based on the long lifetimes of low-mass stars and the requirement that the stars had ended their lives before the formation of the Solar System \citep{Nittler97}. The inferred upper limit of some 2 solar masses almost certainly  reflects chemistry. Above this limit, the third dredge-up accompanying thermal pulses in AGB stars enriches the envelope with carbon, increasing the C/O ratio. Because most mass-loss (and hence dust production) occurs during later thermal pulses, AGB stars more massive than  $\approx$2 M$_{\odot}$ are expected to mostly produce carbonaceous dust, not oxides or silicates, as the latter require O$>$C to efficiently form.  A quantitative model of the types and amounts of dust produced by AGB stars as a function of mass and metallicity has been presented  by \citet{Zhukovska2008}. These authors indeed find a sharp drop-off in the expected production of silicates by AGB stars around 2 solar masses, in good agreement with our inferences from the grain data.  However, they also predict a very large production of O-rich dust in AGB stars of intermediate mass (IM, 4--7 M$_{\odot}$), since  efficient hot-bottom burning (HBB) in such stars maintains a C/O ratio  lower than unity. This is in clear disagreement with the grain data, for which there is no evidence of O-rich  grains from IM-AGB stars. In fact, \citet{Lugaro2007} argued on the basis of unusual Mg isotopes that an unusual Group 2  presolar MgAl$_{\rm 2}$O$_{\rm 4}$ grain, OC2, originated in an IM-AGB star. However, subsequent refinement of the \ox{16}(p,$\gamma$)\fluor{17} reaction rate \citep{Iliadis2008} results in a lower limit of \ox{17}/\ox{16}$\approx$0.002 from hot-bottom burning,  convincingly ruling out IM-AGB stars as the sources of any known presolar O-rich grains (Fig.~1). This paucity of grains from IM-AGB stars is puzzling and deserves further attention. We note, however, that the model of \citet{Zhukovska2008} likely greatly overestimates the level of HBB in solar-metallicity stars. These authors assumed that HBB is efficient in all IM-AGB stars, but stellar evolution models predict the efficiency of HBB to decrease with increasing metallicity. For example, the FRANEC model does not predict HBB in 5M$_{\odot}$ AGB stars of solar metallicity \citep{Zinner2006}.

\section{Metallicity Distributions and GCE}

\begin{figure}[h!]
\begin{center}
\includegraphics[scale=.65, angle=0]{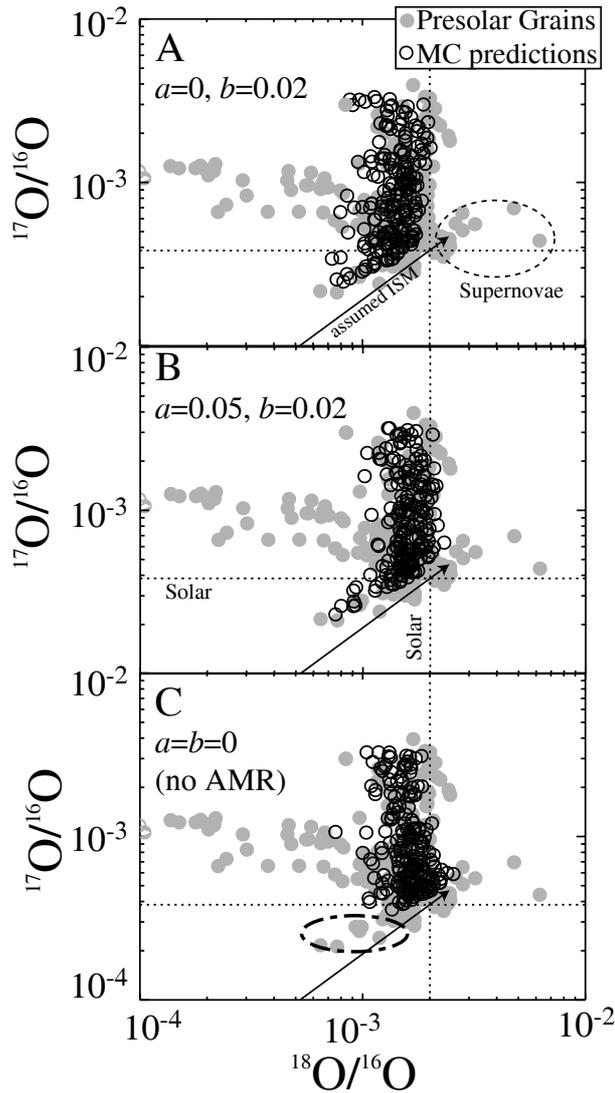}
\caption{Comparisons of Monte Carlo predictions (parameters indicated) with observed O-isotopic ratios in presolar grains. Initial O isotopic ratios of stars are assumed to have Solar \ox{17}/\ox{18} ratios (``assumed ISM'').  Indicated \ox{18}-rich grains likely formed in supernovae and are not expected to be explained by Monte Carlo model of AGB star populations. Models in A and B  assume the existence of an age-metallicity relationship (AMR) in the Milky Way disk; that in C does not. See text for details.}\label{mcplots1}
\end{center}
\end{figure}
\begin{figure}
\begin{center}
\includegraphics[scale=.45, angle=0]{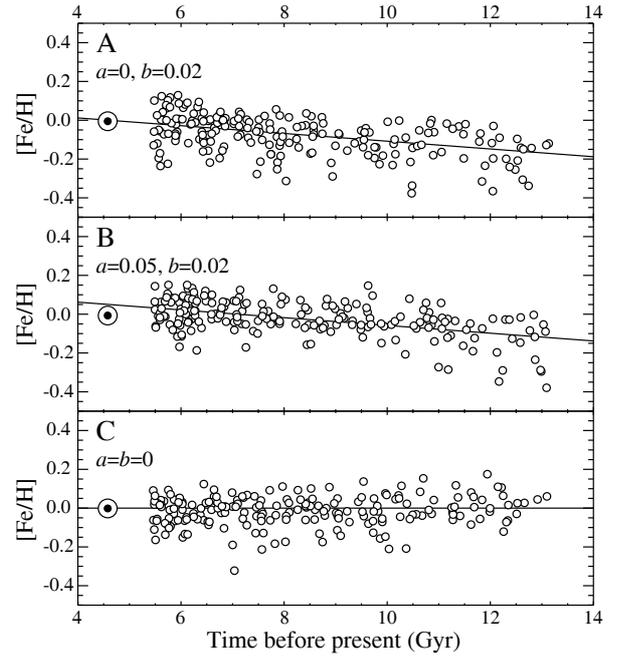}
\caption{Metallicities and formation times of simulated stars in Monte Carlo models corresponding to Fig.~3. Average age-metallicity relations assumed for each case are shown as solid lines. }\label{feohvt}
\end{center}
\end{figure}

Simulated O isotopic distributions, corresponding to three sets of Monte Carlo  parameters, are compared to the grain data in Figure \ref{mcplots1}. Corresponding metallicities and formation times for the model stars are shown in Figure \ref{feohvt}. Note that the goal is to match the distribution of Group 1 and 3 grains (Fig.~1); the model does not include any extra-mixing as needed to explain the \ox{18} depletions in Group 2 grains. All models assume the mass range described in the previous section (1.15--2.2 M$_{\odot}$). In Fig.~\ref{mcplots1}A, we assume that the O isotopic ratios vary linearly with metallicity Fe/H (the same assumption used by BS99) and the AMR parameters are taken to be $a=0$ and $b=0.02$ dex/Gyr, with a metallicity scatter of $\sigma_{Z}$=12\% at solar metallicity. The predicted distribution is slightly shifted to lower \ox{18}/\ox{16} ratios than the observed distribution of grains as evidenced by the significant number of \ox{17}-rich grains along the solar  \ox{18}/\ox{16} axis not matched by the model. This reflects the fact that the average \ox{18}/\ox{16} ratio of Group 1 grains is well-explained by the first dredge-up in stars of initially solar O-isotopic composition. However, the parent stars are older than the Sun and thus are assumed in the model to have, on average, lower-than-solar initial \ox{18}/\ox{16} ratios.  In order to reproduce the data,  the Sun must be anomalous for its time of formation in either metallicity or O-isotopic composition. Fig.~\ref{mcplots1}B shows results for the same model parameters as in panel A, except that the parameter $a$ is set to 0.05 dex. That is, it is assumed that at the time of Solar birth, the average metallicity of the local disk was [Fe/H]=0.05, about 11\% higher than Solar. This model provides a remarkably good match to the Group 1 and 3 grain distribution, in particular reproducing the trend of the \ox{16}-rich Group 3 grains. An almost identical distribution is found if $a$ is set to zero, but the initial O isotope-metallicity relationship is modified such that the Sun is depleted in \ox{16} by about 11\% for its metallicity.  

For the simulations in Fig.~\ref{mcplots1}A and B, the  $\sigma_{Z}$ parameter was adjusted to match (by eye) the width of the distribution. The resulting spread in [Fe/H] values (Fig.~\ref{feohvt}A and B) ranges from $\approx$0.06 dex to $\approx$0.09 dex. These values are  smaller than many astronomical estimates of the intrinsic metallicity scatter for stars born at the same time (e.g., $\sigma=0.2$ dex, Holmberg et al. 2007)\nocite{Holmberg2007}, but are consistent with the very small metallicity dispersion measured in the local ISM \citep{Cartledge2006}. However, a larger metallicity dispersion could be accommodated  by different assumptions regarding the O isotope-metallicity relationship.

Several authors \citep{Feltzing2001,Nordstrom2004,Holmberg2007}  have questioned the existence of an AMR in the Galactic disk, based on data from the Hipparcos satellite. To investigate this issue, we calculated Monte Carlo simulations under the assumption of no AMR ($a=b=0$); results are shown in  Fig.~\ref{mcplots1}C. The Group 1 grain distribution is reproduced reasonably well in this case, but no stars are predicted with the sub-solar \ox{17}/\ox{16} and \ox{18}/\ox{16} ratios observed in the Group 3 grains (dash-dot ellipse).  We have not identified model parameters that can reproduce the general shape of the Group 1 and 3 O isotope distribution without including an AMR to explain the Group 3 tail. The much better agreement for simulations which include an AMR (e.g. Fig.~\ref{mcplots1}B) compared to models without an AMR strongly argues that an AMR did exist in the Milky Way disk prior to the formation of the Sun 4.6 Gyr ago.  

\section{The Solar \ox{17}/\ox{18} ratio.} 

It has been recognized for many years that molecular clouds throughout the Galaxy  have a uniform  \ox{17}/\ox{18} ratio that is distinctly higher than that of the Solar System ($\approx$0.25--0.3 compared to 0.19; Penzias 1981; Wilson \& Rood 1994; Wouterloot et al. 2008 )\nocite{Penzias81,Wilson94,Wouterloot2008}. Recent observations of protostars have found similarly high \ox{17}/\ox{18} ratios \citep{Young2008}. The origin of this discrepancy is unknown but one favored explanation is that the apparently atypical Solar ratio is due to some sort of ``local'' event associated with its parental molecular cloud. For example, \citet{Prantzos96} considered a model wherein self-pollution of the  cloud by a generation of massive stars decreased the initial \ox{17}/\ox{18} ratio to the Solar value. Young et al. (2008) propose a very similar scenario.  However, in previous work we have shown that the presolar grain data, especially those of the Group 3 grains,  are well-explained by an origin in AGB stars with Solar-like initial \ox{17}/\ox{18} ratios \citep{Nittler97b}. In particular, since the first dredge-up can only increase the surface \ox{17}/\ox{18} ratio of a star and many presolar grains have values for this ratio lower than the present-day interstellar value, at least some stars going back billions of years before the formation of the Sun must have had low initial \ox{17}/\ox{18} ratios. We can use the Monte Carlo simulations to further investigate this issue.
\begin{figure}[h]
\begin{center}
\includegraphics[scale=.65, angle=0]{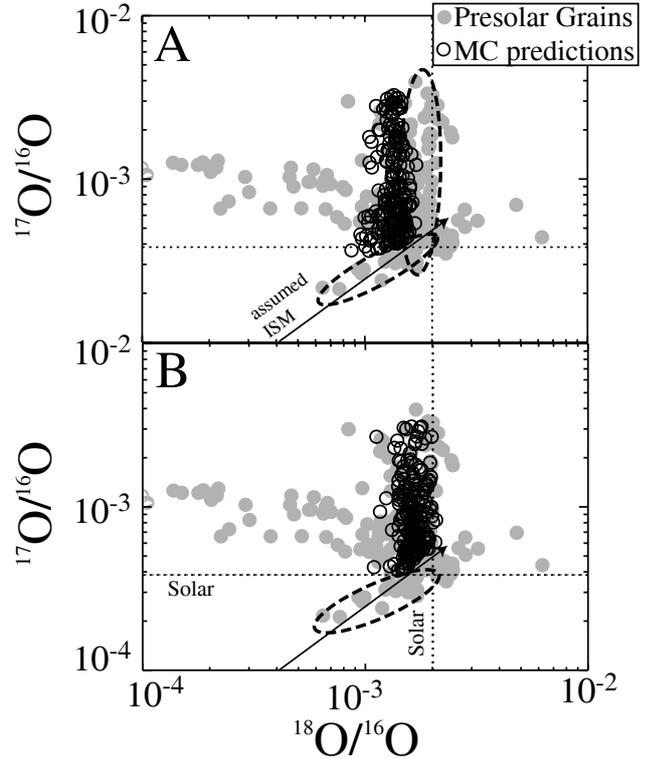}
\caption{Comparisons of Monte Carlo predictions (open circles) with observed O-isotopic ratios in presolar grains (filled circles). Initial O isotopic ratios of stars are assumed to have \ox{17}/\ox{18}=0.25, similar to observations of the present-day ISM (``assumed ISM''). A) Model assuming Sun has unusual \ox{18}/\ox{16} ratio for its age. B) Model assuming Sun has unusual \ox{17}/\ox{16} ratio for its age.  See text for details.}\label{mcplots2}
\end{center}
\end{figure}

In Fig.~\ref{mcplots2}, we show  results of Monte Carlo models in which the initial compositions of the simulated stars are assumed to have \ox{17}/\ox{18} ratios similar to those observed in the present-day interstellar medium (``assumed ISM'' in the Figure). Other than the assumed O isotope evolution, the model parameters are the same as the one shown in Fig.~\ref{mcplots1}B. For Fig.~\ref{mcplots2}A, it was assumed that the Solar \ox{17}/\ox{16} ratio is typical for its metallicity, but the Solar \ox{18}/\ox{16} ratio is greatly enhanced (Prantzos et al. 1996). In this case, the model clearly fails to match the data, missing essentially all  of the Group 3 grains and most Group 1 grains (dashed ellipses). One can get a slightly better overlap of the predictions with the grain data with a very high assumed value of the GCE $a$ parameter, $\approx$0.3 dex. However, even in this case, the fit is relatively poor and the grains plotting below the assumed ISM line (i.e., most Group 3 grains) cannot be explained. Moreover, this value of $a$ would indicate that typical stars of the Sun's age have metallicity higher than the Sun by a factor of two, which is not consistent with any observations (e.g., Reddy et al. 2003)\nocite{Reddy2003}. Fig.~\ref{mcplots2}B shows the case where  the Solar \ox{18}/\ox{16} ratio is assumed to be typical, but its  \ox{17}/\ox{16} ratio is unusually low. This model can match the Group 1 data reasonably well but  completely misses the Group 3 grains (ellipse) as well as Group 1 grains plotting below the assumed ISM line. 

All in all, the very good agreement with the data for models that assume a Solar \ox{17}/\ox{18} ratio for the Galaxy (e.g. Fig.~\ref{mcplots1}B) compared to those that assume a significantly higher ratio (Fig.~\ref{mcplots2}) argues that the Solar  ratio was not atypical for the presolar Galaxy and is not due to a ``local'' (e.g., molecular cloud level) event.  

Of course, if the Sun's composition was affected by self-pollution by supernova ejecta in a molecular cloud, the same process might affect a significant fraction of other stars as well. Thus, it is possible that the observed present-day interstellar ratio was typical in presolar times as well and the grains with lower initial \ox{17}/\ox{18} ratios formed in stars that also experienced such pollution. This needs to be quantitatively modeled. However, it is not clear how such a process could yield the distinctive ``tail'' of the Group 3 grain O-isotope distribution, a tail that is naturally explained by assuming Solar \ox{17}/\ox{18} ratios for low-mass stars born billions of years before the Sun (Fig.~\ref{mcplots1}B).

\section{Conclusions}

We have used simple Monte Carlo techniques together with predictions of the first dredge-up in red giant stars to simulate the  distribution of O isotopes in populations of  AGB stars that might have provided presolar stardust grains to the protosolar cloud. The models are oversimplified, yet allow for some general conclusions:

1) The distribution of \ox{17}/\ox{16} ratios in presolar grains indicates that the parent stars had a mass distribution roughly following a Salpeter (1955) initial mass function, with a mass range of  $\sim$1.15 -- 2.2 M$_{\odot}$. The lower mass cutoff corresponds to the lowest mass that would have had time to evolve to the AGB phase by the time of solar System formation 4.6 Gyr ago. The upper mass cutoff reflects the fact that above this mass, AGB stars efficiently become C stars, producing C-rich dust rather than O-rich dust. However, hot-bottom burning (HBB)  in intermediate-mass ($>$4M$_{\odot}$) AGB stars is expected to maintain O$>$C and lead to a large production of O-rich dust \citep{Zhukovska2008}. Such grains are not observed among the presolar grain population. This discrepancy is not understood, but it is likely that HBB in solar-metallicity stars is overestimated in the models of  \citet{Zhukovska2008}.

2) The O-isotopic data of presolar stardust indicate that an average age-metallicity relationship existed in the presolar galactic disk. 

3) Explaining the distribution of O-isotopic ratios of presolar stardust with our model requires that the Sun has either a slightly lower metallicity for its age (but within the expected scatter for stars of a given age) or slightly unusual O isotopic ratios for its metallicity, or both.

4)  The Solar \ox{17}/\ox{18} ratio, while significantly lower than that observed in present-day molecular clouds and protostars, was not atypical for the presolar solar neighborhood. In particular, the low-mass stellar parents of Group 3 grains, which  must have formed billions of years prior to the Sun, must have had Solar-like  initial  \ox{17}/\ox{18} ratios. This argues against self-pollution of the protosolar molecular cloud by supernova ejecta as an explanation for the unusual Solar ratio. A satisfying explanation for this O-isotope puzzle is still lacking, but a presolar galactic merger event may have played a role \citep{Clayton2003,Clayton2004b}.

\section*{Acknowledgments} 

I would like to thank Roberto Gallino for many years of friendship and scientific collaboration and especially for his tireless championing of presolar grains as useful tools for astronomy. This paper benefited from useful conversations with Roberto Gallino and Maria Lugaro.





\end{document}